# Occupational Safety within Non-Routine Manufacturing Processes: Evaluating the Validity of Task-Based Ergonomic Assessments

Charu Tripathi*[a], Manish Arora[a], and Amaresh Chakrabarti[a]

[a] *Department of Design and Manufacturing (DM), Indian Institute of Science (IISc), Bengaluru (560012), India*

*Corresponding author: charut@iisc.ac.in

# Occupational Safety within Non-Routine Manufacturing Processes: Evaluating the Validity of Task-Based Ergonomic Assessments

**Abstract**

Direct measurement ergonomic assessment is reshaping occupational safety by facilitating highly reliable risk estimation. Industry 5.0, advocating human-centricity, has catalysed increasing adoption of direct measurement tools in manufacturing industries. However, due to technical and feasibility constraints in their practical implementations, especially within non-routine manufacturing processes, task-based approach to ergonomic assessment is utilized. Despite enabling operationalization of robust ergonomic assessment technologies within complicated industrial processes, task-based approach raises several validity concerns. Hence, to ascertain functional utility of the resultant safety interventions, this study evaluates the construct validity of task-based ergonomic assessment within non-routine work utilizing Multitrait-multimethod (MTMM) matrix followed by video-based content analysis. Ergonomic exposure traits were collected for 46 participants through direct measurement and self-reported techniques utilizing inertial motion capture and Borg's RPE rating scale respectively. Findings include unsubstantiated convergent validity (low same-trait correlations: 0.149-0.243) and weak evidence of discriminant validity with statistical significance ($p<0.001$). The study also identifies three primary factors undermining construct validity through video-based content analysis. Findings also elucidate misinterpretation of ergonomic risk and action levels. Therefore, practical implications entail underestimation of actual ergonomic risks when estimated through task-based assessment. This highlights the need for enhancement in ergonomic assessment technologies focused on cumulative load analysis compatible within diverse industrial processes.

Keywords: Direct measurement ergonomic assessment, Inertial motion capture, Occupational safety, Non-routine manufacturing processes, Task-based ergonomics

## 1. Introduction

Occupational safety is of utmost importance within manufacturing industries due to the inherently strenuous nature of a majority of tasks. Throughout the years, the significance of ergonomics within manufacturing industries has become progressively evident, driven by the imperative to ensure occupational safety, enhance productivity, and optimize operational efficiencies [1–6]. With the emergence of Industry 5.0, the focus of manufacturing process planning is shifting from purely efficiency-driven approach to a more human-centric approach [7–10]. The manufacturing sector, poses a plethora of risk factors such as repetitive motion, force exertion, heavy loads, awkward or stationary postures, excessive vibration etc. to the workers which leads to work-related musculoskeletal disorders (WMSDs) [11–13]. Manufacturing processes involve diverse working scenarios varying in terms of constituent tasks, repeating or cyclic patterns of tasks, controllable and uncontrollable disturbances at shop-floor and so on [14–17]. While accurate and precise risk estimation within every working scenario remains an indispensable prerequisite for occupational safety, undertaking effective ergonomic assessment in each working scenario poses its own specific challenges.

With technological advancements, the ergonomic assessment strategies have been refined to accommodate various working scenarios whilst utilising more sophisticated techniques for risk estimation. In order to ascertain the reliability and accuracy of the ergonomic assessments, there has been increasing adoption of advanced industrial engineering technologies such as direct measurement (DM) tools [18–20]. The DM tools involve quantification of exposure parameters through sensors mounted on the relevant body segments, for example, motion capture systems for posture identification, sEMG for muscle activity analysis and EEG for cognitive load assessments. However, due to technical and feasibility constraints posed by manufacturing shop floors, DM tools require a limited sample of work which can be considered to represent the ergonomic aspects of a complete work shift, and the sample exposure is weighted to estimate the overall exposure throughout a work shift. This approach is easily applicable to routine work, considering a few work cycles as the representative sample for ergonomic assessment, owing to relatively short cycle times.

However, DM ergonomic assessment in non-routine manufacturing processes becomes markedly complicated due to the absence of a repeating segment of work. The variation in both sequence and content of the underlying tasks within non-routine manufacturing processes omit any representative work sample to be utilized for ergonomic assessments [14–16]. Manual material handling (MMH) presents a good example of non-routine work. MMH activities involve variability in sequence and content, lower degree of standardisation in operating procedures, and dynamic decisions among other factors that introduce unstructuredness in the process flow [21–23]. Despite the complexity in the assessment process, the nature of activities demands highly reliable and accurate ergonomic risk estimation since MMH activities are physically taxing, and thereby, substantially impact the safety considerations. In case of non-routine work and other complex working scenarios, due to the absence of such a work sample, task-based ergonomic assessment approach provides a feasible and practical solution [24].

In the task-based approach, the details regarding all the activities scheduled in a work shift are collated through process flow SOPs, structured interviews, work logs, surveys etc. to shortlist ergonomically critical activities [24,25]. These critical activities are separately utilised for ergonomic assessment to calculate the exposure scores instead of the complete process flow to comply with the feasibility constraints in data collection. The bypassed work elements, such as the projected time duration (both continuous or intermittent) of these activities in a complete work shift, the nature of loading (dynamic or static loading), postural support available to the workers and so on, are taken into account with the application of correction factors on the calculated exposure scores to estimate the overall risk. On the contrary, an assessment involving continuous observation throughout the complete work shift is referred to as cumulative load assessment [21,26].

Even though the task-based approach is a very commonly practiced, feasible solution for application of DM ergonomic assessment in non-routine, complex or highly dynamic work, there has been theoretical concerns regarding validity of results [15,27,28]. This is attributable to the elevated probability that multiple work elements get overlooked, considering the assessment is focused on a subset of the work, with the remaining elements being addressed through approximation. To attain effective ergonomic risk estimation it is essential to address any potential research-practice gap prevalent in assessment methodologies [29–33]. Stanton and Young [33] have argued that reliability alone is not sufficient and an extra effort to ensure validity of measurements is critical in ergonomics as methods can be reliable without being valid. Hence, the DM ergonomic assessments conducted through task-based approach, although delivering reliable estimates, makes it worthwhile to evaluate whether the estimated risks realistically replicate the actual ergonomic exposure.

In epidemiological research, validity concept has been broadly classified into four types i.e. face validity, construct validity, content validity, and criterion validity. Chronologically, face validity is the preliminary type that pertains to the extent to which an ergonomic assessment approach appears, at face value, to measure what it claims to measure, as judged by non-experts or participants [34]. In extension to the face validity, construct validity refers to the degree to which an assessment approach accurately measures the specific theoretical constructs it aims to evaluate [35,36]. Subsequently, content validity evaluates the extent to which an ergonomic assessment covers the relevant aspects of those specific constructs [37]. And finally, criterion validity assesses how well the results of an ergonomic assessment correlate with other established methods [38]. Therefore, in this context, task-based approach satisfies face validity as it is extensively accepted to be measuring the intended construct of ergonomic exposure in a work-shift with consideration of ergonomically critical activities. However, the extent to which this approach of separate evaluation of critical activities aligns with the theoretical framework of a holistic ergonomic assessment should be examined through construct validity. Construct validity serves as the cornerstone for establishing the relevance and accuracy of an assessment's measurements in relation to the underlying theoretical constructs it aims to evaluate [36,39,40].

In epidemiology various methods are utilised to evaluate construct validity such as Q-sort, correlation analysis, confirmatory factor analysis (CFA), exploratory factor analysis (EFA) and multitrait-multimethod (MTMM) matrix [40–44]. MTMM matrix is the most utilized means of construct validity evaluation since it derives insights from both convergent validity (similarity within assessment of same traits) and discriminant validity (dissimilarity between assessment of different traits) observed in the assessment of multiple traits utilizing multiple assessment methods [41,45,46].

In the context of task-based approach, ascertaining construct validity is particularly important since this approach is employed to facilitate more reliable technologies for ergonomic assessments within complex and physically demanding environments necessitating robust risk calculation. Yet, a potential discrepancy may emerge due to the fundamental theory of assessment of critical activities in isolation, rather than within the broader context of the entire task [15,27,28,47–49]. Consequently, if such an approach does not provide an authentic representation of workers' actual exposure levels, leveraging advanced industrial engineering technology for its reliability becomes futile, undermining efforts to achieve valid exposure outcomes. Hence, this paper is an effort towards supporting robust ergonomic assessment within challenging industrial processes by exploring construct validity of the task-based approach for ergonomic assessment in non-routine manufacturing tasks utilising multitrait-multimethod (MTMM) matrix.

## 1.1. Relevance to occupational safety

The study is relevant to occupational safety since non-routine manufacturing processes such as MMH activities are known to culminate into various work-related musculoskeletal disorders (WMSDs) such as carpal tunnel syndrome (CTS), lower back and rotator cuff injuries etc. in absence of effective ergonomic interventions. For ensuring the effectiveness of interventions, the validity of assessments in addition to reliability forms a significant precursor. Hence, with rapid advancements in industrial engineering technologies and complexities in manufacturing processes, it becomes necessary to revisit the validity concerns regarding practical measurements in ergonomics, especially for assessment approaches utilizing room for approximations. Addressing this research-practice gap is significant to ensure risk estimations from the measurement approaches in conjunction with the novel tools are trustworthy for making important decisions regarding manufacturing process planning and workplace design to prevent critical occupational safety issues.

## 2. Literature survey

Practical implementation approaches such as task-based assessment of ergonomic parameters, in order to accommodate feasibility constraints of the advanced assessment tools, often work with data collection for significantly limited time duration, number of subjects etc. Moreover, in several cases task-based assessment is conducted in simulated environment to avoid the challenges posed by highly dynamic manufacturing floors, along with disturbances such as electromagnetic interference (EMI), mechanical vibrations, acoustic noise, thermal fluctuations, and airborne particulates that can compromise sensor accuracy and data integrity. Subsequently, the findings of constrained ergonomic assessments go through various approximations to become generalizable to the broader set of population, actual process flow and real work environment. However, effective evaluation of ergonomics in epidemiology depends on a lot of complexities and becomes more challenging as room for approximation increases. For instance, Granata et al. [22] talk about how individuals performing same task can still differ in experiencing the same loads and van Dieën et al. [50] show that different people react differently to the resulting ergonomic interventions which is crucial to consider when the assessment involves fewer subjects. Kanis [51] argue that measurements of human performance can accumulate dispersion due to human characteristics such as getting bored, tired or gaining knowledge and experience etc. and studied the dispersion patterning to limit test-retest trials in experimental design. Similarly, Dempsey and Mathiassen [15] present a detailed description of evaluation of manufacturing processes and hence the methods of studying these processes. This detailed account very efficiently describes the various types of task analysis and other approximations specific to non-routine work. Concerns regarding the validity of some approximations have also been discussed concluding with the need for deeper explorations. Dempsey [21] elaborates the crucial correction factors to consider when applying task-based approach specifically in MMH activities regarding time duration and frequency of critical activities. Kumar [52] explains mode of MSDs and establishes that lower back pain can be caused by prolonged history of varied exposures. It is therefore not solely the peak stresses that are of significance which are typically the focus in task-based estimates. Svendesen et al. [53] compares task-based ergonomic exposure estimates with estimates based on occupation and the effect of increasing number of subjects on the precision of task-based estimates. The study concludes task-based estimates as marginally better than occupation based estimates but with insignificant increase in precision with increase in number of subjects.

On a tangential note, the debate regarding the significance of reliability and validity of measurements in ergonomics had received ample attention in past years. Stanton and Young [33] presented an account on the importance of reliable and valid ergonomic findings for overall profit to a firm. Annett [54] proposed a dichotomy of ergonomic studies, classified into analytical and evaluatory, and proposed the validity requirements such as construct and predictive validity corresponding to each type. Similarly, several studies presented discourse on the importance of maintaining reliability and validity of ergonomic research results in practical realm by revisiting validity concerns in light of advancements in ergonomics [55–58]. However, since 2016, there has been a notable decline in academic research on the validity front for the fundamentals of measurement approaches in ergonomics. With the advent of dramatic technological developments, the adoption of high end measurement techniques and increasing complexity in manual tasks mandates revisiting the basic principles such as validity assurance for measurement approaches in ergonomics in order to attain concrete results in practice.

On similar lines, Vigoroso et al. [59] highlight the need for clearer methodologies within assessment approaches under human factors and ergonomics to better integrate safety in industrial practices. Hence, with the availability of very few research studies on the theoretical considerations involved in the approximations required by the task-based approach, and yet widespread practical implementations in order to feasibly incorporate reliability enhancing data collection tools, a crucial research-practice gap was identified.

Addressing WMSDs requires not only accurate measurement of relevant ergonomic risk parameters but also the use of an assessment approach that ensures the alignment of the outcome with the intended constructs, thereby establishing construct validity. To the best of the authors' knowledge, no empirical study has been conducted for studying the construct validity of task-based approach. Hence, this study explores the construct validity of task-based ergonomic assessment approach using two case studies of non-routine manufacturing tasks.

## 3. Methodology

This study evaluates the construct validity of task-based ergonomic assessment of non-routine work using Multitrait-multimethod (MTMM) matrix in conjunction with direct measurement using inertial motion capture technique as well as self-reported method using Borg's RPE rating scale. The methodology followed for construct validity evaluation and follow-up analysis is depicted in Figure 1.

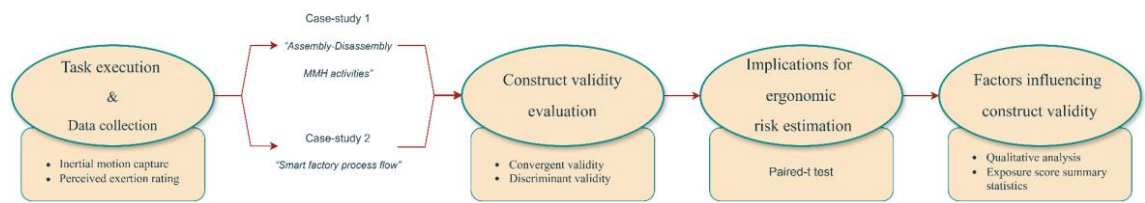

Fig.1. Study Methodology

The MTMM matrix involves measurement of multiple traits using multiple methods in order to ascertain convergent and discriminant validity of the method or approach in question. For the ergonomic assessment, two traits were measured 1) MSD risk score and 2) perceived exertion using two methods 1) task-based assessment and 2) cumulative load assessment.

Xsens MVN Awinda motion capture system was used as the direct measurement tool to calculate MSD risk score and Borg's RPE self-reporting scale was used to quantify the perceived exertion in two case studies.

### *3.1. Participant Selection and Task Design*

The experiment was structured around two distinct case studies to comprehensively explore the construct validity of task-based approach using different process flows and work environments. The experimental protocol was approved by the institutional ethics committee (approval number: 6-31082018) and informed written consent was obtained from the participants prior to experimentation.

### *3.1.1. Case Study – 1*

The initial case study involved the collection of data from a group of 36 individuals (20 males and 16 females) in the age range of 23 – 32 years, not directly associated with factory work, in a laboratory simulated manufacturing task. This subject selection, although driven by availability, additionally helped in the inclusion of various aspects of non-routine work such as unstructuredness due to the participants' lack of professional expertise, leading to elemental actions executed in varied, non-standardized ways, variable sequence of activities, dynamic decisions, occasional omissions or missteps, further contributing to the variability in task performance.

The task involved assembly and disassembly of an aircraft wingbox along with MMH activities including lifting, lowering, turning and carrying of structural components such as spar cap and web panels as shown in Figure 2(a). The participants were provided with the content and tools to execute the activities involved in the task and they were free to follow any sequence of activities.

*3.1.2. Case Study – 2*

To validate and contextualize the findings from the first case study, subsequent investigation targeted professional manufacturing factory workers. Data collection occurred within the CEFC smart factory, ensuring a direct observation from an actual shop floor.

The observed sample of work involved various activities in the process flow of pneumatic cylinder assembly in the smart factory such as order placement on the Manufacturing Execution System (MES), disassembly of required parts, kitting operations, AGV and robot monitoring, control and troubleshooting, quality inspection and so on as shown in Figure 2(b).

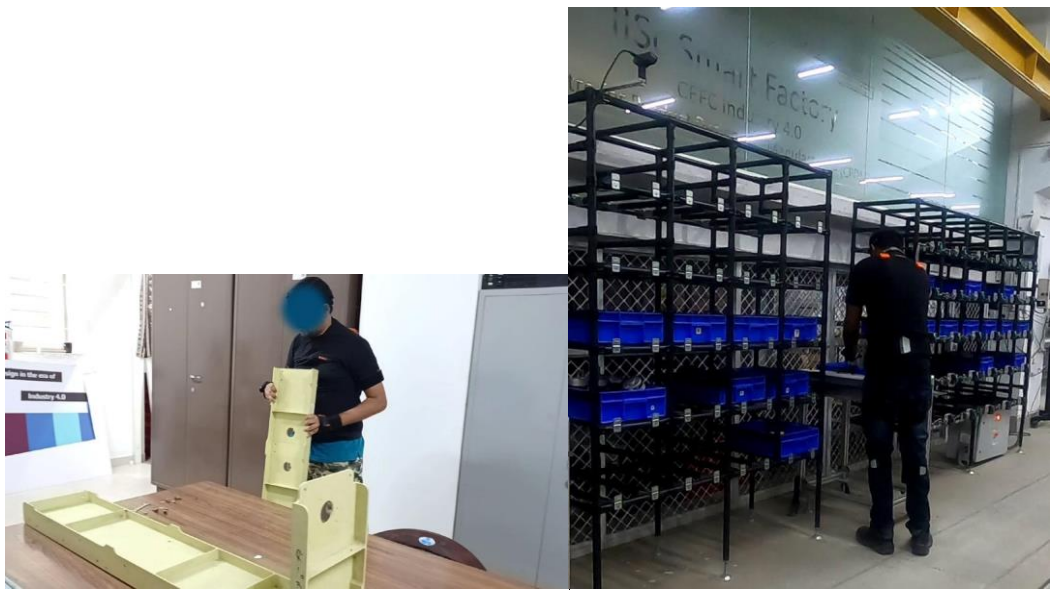

Fig.2. Manufacturing task execution (a) Case study-1, (b) Case study-2

*3.2. Data collection*

In order to observe task-based assessment in conjunction with both direct measurement as well as self-reporting tools, the following data was collected:

(1) Direct measurement: Inertial motion capture using MVN Awinda MoCap system to obtain posture specific data during the task execution.

(2) Self-report: Borg's RPE scale was used after task completion to quantify perceived exertion.

Data collection was carried out in two distinct sessions for each participant for ergonomic assessment using cumulative load assessment method and task-based assessment method separately.

*3.2.1. Cumulative load assessment:*

The entire task was observed using both the aforementioned techniques of data collection. Each participant completed the whole process flow donned with the inertial motion capture suit setup and responded to the Borg's RPE survey directly after task completion.

*3.2.2. Task-based assessment:*

Critical activities in both the case studies were identified based on literature survey and industrial best practices regarding similar tasks in the manufacturing shop floors. The participants performed these critical activities with the application of inertial motion capture suit and filled out the Borg's RPE survey after completion of each critical activity.

The postural data collected using inertial motion capture system was used to calculate MSD risk score estimated using Rapid Upper Limb Assessment (RULA) exposure criterion. Table 1 includes the time weighted RULA scores calculated for each participant using methods of cumulative and task-based assessment in both the case studies.

Borg's RPE ratings used to quantify perceived exertion for each participant during methods of cumulative and task-based assessment for both the case studies are also tabulated in Table 1.

*Table 1.* Time weighted average (TWA) RULA scores and mean Borg's perceived exertion ratings

| Case study-1 | Task-based assessment | | Cumulative load assessment | | Case study-2 | Task-based assessment | | Cumulative load assessment | |
|---|---|---|---|---|---|---|---|---|---|
| Participants | Mean RPE | TWA RULA | Mean RPE | TWA RULA | Participants | Mean RPE | TWA RULA | Mean RPE | TWA RULA |
| P1 | 10.8 | 4.46 | 13.5 | 4.20 | P1 | 12.0 | 4.06 | 15.7 | 4.72 |
| P2 | 10.0 | 3.67 | 13.0 | 4.00 | P2 | 11.7 | 4.06 | 15.0 | 4.90 |
| P3 | 11.0 | 4.12 | 16.0 | 5.85 | P3 | 11.3 | 4.06 | 15.0 | 5.02 |
| P4 | 10.8 | 3.71 | 14.0 | 5.01 | P4 | 12.0 | 3.58 | 15.0 | 4.68 |
| P5 | 10.5 | 3.81 | 13.3 | 5.18 | P5 | 12.0 | 4.09 | 15.7 | 4.39 |
| P6 | 10.0 | 3.47 | 14.5 | 5.16 | P6 | 11.3 | 4.08 | 15.7 | 5.29 |
| P7 | 11.3 | 4.45 | 13.0 | 5.00 | P7 | 12.0 | 3.52 | 15.0 | 5.03 |
| P8 | 11.3 | 4.59 | 14.0 | 5.15 | P8 | 11.3 | 4.06 | 14.7 | 4.01 |
| P9 | 10.8 | 4.76 | 13.5 | 4.23 | P9 | 10.0 | 4.09 | 15.0 | 4.42 |
| P10 | 11.0 | 4.98 | 13.8 | 4.86 | P10 | 12.0 | 3.50 | 14.7 | 4.69 |
| P11 | 10.5 | 4.27 | 13.5 | 3.96 | | | | | |
| P12 | 10.8 | 4.51 | 14.3 | 5.36 | | | | | |
| P13 | 10.8 | 4.17 | 12.8 | 3.76 | | | | | |
| P14 | 10.5 | 4.53 | 14.0 | 4.05 | | | | | |
| P15 | 10.5 | 4.60 | 13.3 | 5.31 | | | | | |
| P16 | 11.0 | 4.52 | 13.8 | 4.75 | | | | | |
| P17 | 10.3 | 3.87 | 13.5 | 3.88 | | | | | |
| P18 | 10.8 | 4.32 | 13.8 | 4.96 | | | | | |
| P19 | 10.5 | 4.44 | 13.8 | 4.10 | | | | | |
| P20 | 10.5 | 4.33 | 13.8 | 4.01 | | | | | |
| P21 | 10.8 | 4.64 | 14.0 | 5.16 | | | | | |
| P22 | 10.8 | 4.47 | 13.8 | 5.31 | | | | | |
| P23 | 11.3 | 4.49 | 13.8 | 5.23 | | | | | |
| P24 | 10.3 | 4.42 | 14.0 | 4.20 | | | | | |
| P25 | 10.8 | 4.66 | 13.8 | 5.20 | | | | | |
| P26 | 10.8 | 4.52 | 13.8 | 5.31 | | | | | |
| P27 | 11.0 | 4.38 | 12.5 | 3.99 | | | | | |
| P28 | 11.0 | 4.33 | 13.3 | 5.24 | | | | | |
| P29 | 10.5 | 4.39 | 14.0 | 4.02 | | | | | |
| P30 | 10.3 | 4.38 | 12.8 | 4.17 | | | | | |

| P31 | 11.0 | 5.62 | 14.3 | 5.30 |
| P32 | 10.5 | 4.03 | 13.8 | 3.94 |
| P33 | 10.8 | 4.45 | 14.3 | 4.04 |
| P34 | 11.0 | 4.29 | 13.5 | 3.99 |
| P35 | 10.3 | 3.98 | 12.8 | 3.92 |
| P36 | 11.0 | 3.98 | 14.0 | 3.91 |

### *3.3. Data analyses*

#### *3.3.1. Multi-trait Multi-method matrix*

The statistical analysis for exploring construct validity was done using MTMM matrix. MTMM matrix was chosen as it supports the evaluation of both convergent as well as discriminant validity by examining the inter-relationship between various traits calculated using the different methods.

In this study, the two methods being compared are cumulative load assessment and task-based assessment. Two traits calculated using the collected data are MSD risk score using postural data from the motion capture system and perceived exertion quantified using self-reported Borg's scale.

The following four types of inter-relations were quantified using Pearson's correlation coefficients and p values in order to form the MTMM matrix:

(1) Same traits calculated using same methods

(2) Same traits calculated using different methods

(3) Different traits calculated using same methods, and

(4) Different traits calculated using different methods

The correlation coefficients for both the case studies are tabulated in the form of MTMM matrices in Table 2(a) and Table 2(b).

3.3.1.1. Interpretation of MTMM matrix

Campbell and Fisk [41] developed Multitrait-Multimethod (MTMM) Matrix, a tool for evaluation of construct validity in applied research through assessment of convergent and discriminant validity. Convergent validity provides the extent to which theoretically similar traits are empirically interrelated. Similarly, discriminant validity provides the extent to which theoretically dissimilar traits are empirically not interrelated. Subsequently, Campbell and Fisk [41] put forward the following main criteria for the interpretation of MTMM matrix to ascertain convergent and discriminant validity for any method being studied.

(1) The coefficients in reliability diagonal should be consistently the highest in the MTMM matrix. This ensures that the individual traits are internally consistent.

(2) The coefficients in validity diagonal should be significantly different from zero preferably medium to high strength of correlation. This ensures same traits are correlated sufficiently and forms evidence of convergent validity.

(3) In a heteromethod block, validity coefficient should be higher than all other values in its corresponding row. This ensures that same traits measured through different methods are correlated stronger than different traits measured through different methods and forms evidence of discriminant validity.

(4) Similarly, in a heteromethod block, validity coefficient should be higher than all other values in its corresponding column. This ensures that different traits measured through same methods even are not as strongly correlated as same traits measured through different methods and forms evidence for discriminant validity.

(5) A validity coefficient should be higher than all coefficients in the heterotrait-monomethod triangles. This essentially emphasizes that trait factors should be stronger than methods factors.

*3.3.2. Paired t-test:*

Paired t –test was conducted for more clarity on the similarity or difference in the ergonomic exposures calculated using task-based and cumulative methods beyond correlations (i.e. strength and direction of relationships) by comparison of the mean ergonomic exposures using the 2 methods. This is particularly important in case of ergonomic assessment scores since the organizational decisions on the requirement of ergonomic interventions largely depend on the levels of ergonomic exposures risk rather than absolute risk scores. Hence, similar direction and high strength of relationship alone are not sufficient but the exposure scores calculated using task-based as well as cumulative method should fall in the same action levels as per the ergonomic assessment criteria in order to lead to similar decisions on intervention requirement in the workplace which can be ensured by comparison of mean exposures.

*3.3.3. Qualitative analysis:*

Content analysis involving task execution videos and ergonomic risk data derived from inertial motion capture system was conducted for deeper interpretation of the results of MTMM matrix and t-test by comparing the full spectrum of exposure scores calculated using the 2 methods at a frequency of 60Hz for each participant against the corresponding activities being performed. The activities responsible for mismatch in the exposure scores, the underlying reasons and frequency of occurrences are elaborated in the Discussion section (Section 5).

**4. Results**

*4.1. Construct validity*

The resulting MTMM matrices for both the case studies shown in Table 2 satisfy the 1$^{st}$

criterion of interpretation included in section 3.4.1.1 as all the coefficients in the reliability diagonal are moderate to high in the range of 0.600 to 0.849 for case study–1 and in the range of 0.641 to 0.903 for case study–2. This suggests that the individual traits are internally consistent and suitable for further investigation.

*Table 2.* MTMM matrices for both the case studies (pink cell: reliability coefficients, green cell: validity coefficients i.e. same-trait different-method correlation coefficients)

(a) Case study – 1

| Lab | | Cumulative | | Task-based | |
|---|---|---|---|---|---|
| | | RULA | RPE | RULA | RPE |
| Cumulative | RULA | 0.625114 | | | |
| | RPE | 0.453192 | 0.615296 | | |
| Task-based | RULA | 0.243241 | 0.35383 | 0.849515 | |
| | RPE | 0.38997 | 0.149487 | 0.502063 | 0.600799 |

(b) Case study – 2

| SMFT | | Cumulative | | Task-based | |
|---|---|---|---|---|---|
| | | RULA | RPE | RULA | RPE |
| Cumulative | RULA | 0.903129 | | | |
| | RPE | 0.316007 | 0.63338 | | |
| Task-based | RULA | -0.21176 | 0.457513 | 0.801564 | |
| | RPE | 0.198065 | 0.160393 | -0.4872 | 0.641418 |

*4.1.1. Convergent validity:*

In order to ascertain convergent validity, the inter-relationship between same traits measured using different methods should be strong. Therefore, following inferences regarding convergent validity can be drawn from resulting MTMM matrix referring to the 2nd criterion of interpretation of MTMM matrix from Section 3.3.1.1. From Table 2(a), it is evident that the strength of correlation between same-trait different-method data in case study – 1 is low in the range of (0.149 – 0.243) with statistical significance (p-values < 0.001).

Similarly, from Table 2(b), the strength of correlation between same-trait different-method data regarding case study – 2 is also observed to be low in the range of (-0.211 – 0.160) with statistical significance (p-values < 0.001).

Hence, the Pearson's rho coefficients suggest inadequate convergent validity in both case studies.

*4.1.2. Discriminant validity:*

Discriminant validity can be ascertained if the inter-relationship between same-trait different-method data is stronger than the inter-relationships between all types of different-trait data. Therefore, inferences regarding discriminant validity in both case studies considering the 3$^{rd}$ and 4$^{th}$ criteria of interpretation of MTMM matrix from Section 3.3.1.1, are as follows.

From Table 2(a), in case-study 1, the strength of correlation between same-trait different-method data (0.149 - 0.243) is comparatively lower than the correlation between all different trait data measured using same as well as different method (0.353- 0.502) with statistical significance (p-values < 0.001).

From Table 2(b), in case-study 2 as well, different traits are seen to have stronger correlations whether measured using same or different methods (0.198 – 0.457) compared to same traits measured with different methods (-0.211 – 0.160) with an exception of moderate negative correlation ($\rho$ = -0.487) in different traits measured using task-based method with statistical significance (p-values < 0.001).

Hence, partial satisfaction of discriminant validity criteria with weak correlations in same traits suggests limited evidence of discriminant validity in case study–2, whereas in case study–1, no evidence of discriminant validity could be ascertained.

## 4.2. Paired t-test

Table 3 indicates a statistically significant difference between the TWA RULA scores, with cumulative load scores (M = 4.832, SD = 0.578) higher than scores calculated using task-based method (M = 4.166, SD = 0.377), t(35) = 6.973, p < 0.0001 in case study–1. Similarly, TWA RULA scores increase from (M = 3.909, SD = 0.259) to (M = 4.714, SD = 0.373) when calculated using task-based and cumulative load methods respectively (t(9) = 5.206, p = 0.0006) in case study–2. This suggests that TWA RULA scores when calculated using task-based method may lead to underestimation of the ergonomic risk levels as per RULA criterion. For instance, in case study–1, the TWA RULA scores for 20 out of 36 participants fall in the range of 5 – 6 corresponding to 'medium risk level' using cumulative load calculations all of which fall in the range of 3 – 4 using task-based calculations corresponding to 'low risk level'.

*Table 3*. Paired t statistics for both the case studies

(a) Case study – 1

| Summary statistics | Task-based assessment | Cumulative load assessment |
|---|---|---|
| Mean | 4.166 | 4.832 |
| SD | 0.377 | 0.578 |

| Paired differences | |
|---|---|
| Mean | 0.666 |
| SD | 0.201 |
| t | 6.973 |
| df | 35 |
| p (2-tailed) | < 0.0001 |

(b) Case study – 2

| Summary statistics | Task-based assessment | Cumulative load assessment |
|---|---|---|
| Mean | 3.909 | 4.714 |
| SD | 0.259 | 0.373 |

| Paired differences | |
|---|---|
| Mean | 0.805 |
| *SD* | 0.114 |
| *t* | 5.206 |
| *df* | 9 |
| *p (2-tailed)* | 0.0006 |

## 5. Discussion

In the context of this empirical study, it is noteworthy to mention that the reliability coefficients for each method employed were observed to be moderate to high (0.6 – 0.9). Despite the robust reliability of individual methods, the absence of convergent validity and only moderate evidence of discriminant validity remain significant findings. The high reliability coefficients acknowledge the consistency and repeatability of each method within its specific context. However, inadequacy of convergent and discriminant validity suggests a lack of construct validity and requires deeper understanding of the inter-relationships observed, their strengths and potential reasons.

From qualitative content analysis utilizing task performance videos and ergonomic risk data obtained from motion capture system, the comparison of cumulative load and task-based methods revealed three main factors contributing to this mismatch in exposure estimation.

### *5.1. Unplanned activities overlooked in task-based assessment*

Due to the inherently dynamic nature of non-routine work, the manufacturing operators may encounter unplanned activities that are not accounted for in the predefined process flow arising spontaneously due to emergent issues during the work shift. Since the task-based approach relies on prior identification of ergonomically critical activities from the process flow, there is a strong likelihood of not taking into account, the unplanned activities which may pose ergonomic risks.

Consequently, three such unplanned activities were observed in case study–1, whereas case study–2 projected four unplanned activities. These activities are depicted in

Figure 3, along with the details regarding the percentage of participants encountering these unplanned activities as well as the ergonomic risk levels associated with the corresponding activity execution.

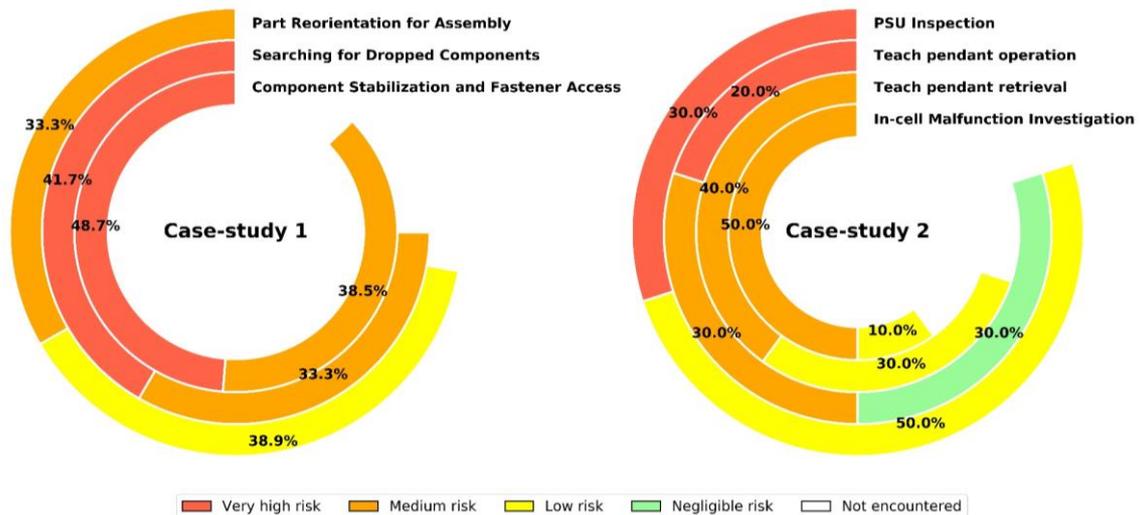

Fig.3. Unplanned activities, percentage of participants and associated RULA scores

For instance, in case study–1, the activities of 'retrieving misplaced components' and 'component stabilisation' posed 'very high risk levels' for 41.7% and 48.7% of the total participants respectively, implying that almost half of the total participants faced high ergonomic risks due to these unplanned activities. While 33.3% and 38.5% of total participants faced medium risk performing the respective activities, whereas 25% and 12.8% participants did not encounter these respective activities.

*5.2. Low-salience activities overlooked in task-based assessment*

Task-based approach entails selection of an optimal set of ergonomically taxing activities considering feasibility constraints in data collection. Consequently, several activities although included in process flow, yet less noticeable or assumed to be not much ergonomically critical, often get overlooked. For instance, Wu and Chung [60] have acknowledged that amongst the MMH activities, 'holding' activity is often overlooked as compared to 'lifting' and 'lowering' activities which are considered primarily ergonomically taxing and empirically ascertained a significant effect of holding mode on the overall perceived ergonomic exposure.

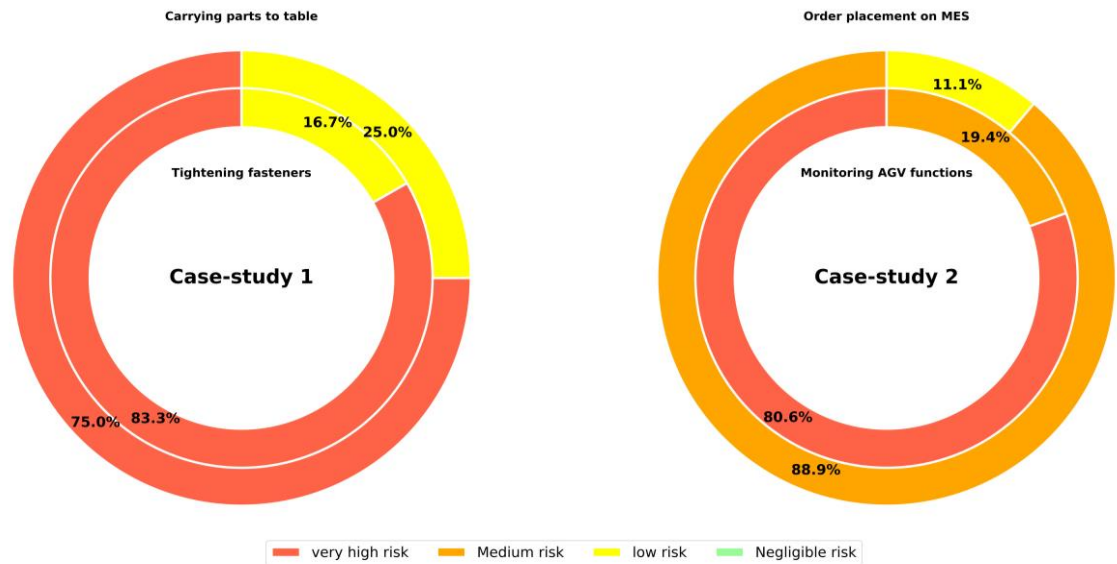

Fig. 4. Overlooked low-salience activities and respective RULA ergonomic risk levels

In the present study, two such low-salience activities have been observed in both the case studies which posed significant ergonomic risks to the participants. These activities are depicted in Figure 4 with details regarding risk levels. For instance, the activity of 'AGV monitoring' involved 'very high risk levels' for 80.6% and 'medium risk levels' for 19.4% of the total participants in case study 2. All four of the identified low-salience activities across both case studies, overlooked in task-based method, being 'carrying', 'tightening fasteners', 'order placement', and 'AGV monitoring', are the kind of activities that lack a degree of standardisation in operating procedure as per postural considerations, and consequently such activities may be generally classified as non-critical from ergonomic assessment standpoint in task-based approach.

*5.3. Same activities resulting in different exposure scores*

In both the case studies, the assessment of same activities (i.e., the selected critical activities) was also observed to culminate in different exposure scores with the utilization of cumulative load and task-based load assessment. Figure 5 depicts the mean and standard deviation of resulting exposure scores for the same activities from both the assessment methods. It is evident from Figure 5 that task-based approach resulted in lower average risk scores in both the case studies. While task-based approach entails execution of the critical activities in isolation, the cumulative load assessment involves the complete process flow involving execution of the same critical activities as a part of an overall goal alongside several intermediate non-critical activities. Consequently, video-based content analysis of the task execution involved in both the assessment methods elucidated two major reasons for lower ergonomic risk scores calculated using task-based method.

(1) Lack of operational context:

In controlled, task-based assessments, activities are performed in isolation, without the context of the overall operational workflow. This abstraction eliminated the broader production goals from the participants' focus and allowed them to focus more on individual activities rather than the overall objective, which in turn heightened their awareness of postural challenges. In contrast, within cumulative load assessment involving the actual process flow, the participants were typically task-oriented, more focused towards the production outcomes, often prioritizing efficiency and throughput over ergonomic self-regulations. This contextual disconnect can be a potential reason for the observed underestimation of ergonomic risks.

(2) Discontinuity in elemental actions:

With the omission of non-critical activities within task-based assessment, the elemental actions connecting the sequential process steps were observed to be modified. Precisely. the discontinuity between elemental actions during inter-activity transitions, stemming from the omission of non-critical activities in task-based assessment, was observed to artificially increase the operational leeway by alleviating some of the postural constraints, inevitable in the actual process flow. This was observed to modify the execution patterns and assumed postures as the same activities were performed in isolation resulting in lower ergonomic risk scores.

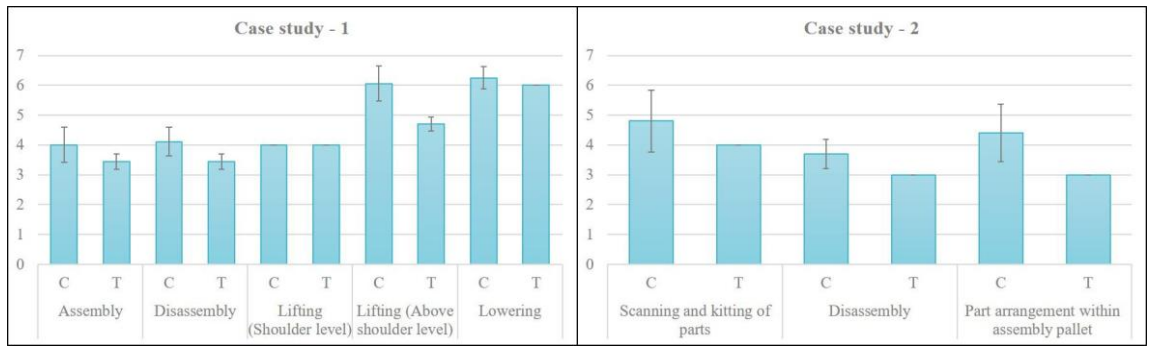

Fig. 5. Mean and standard deviation of critical activities assessed using cumulative load and task-based assessment methods

Consequently, it can be seen from figure 5 that mean RULA scores for almost all the critical activities identified across both the case studies were higher for cumulative load assessment. Moreover, the cumulative load assessment results also show comparatively higher standard deviation (Figure 5) whereas task-based assessment results show zero variance for most of the critical activities. This suggests higher repeatability in task-based assessment, however, it also highlights loss of information from the actual operational scenario as captured in the cumulative load assessment. Also, inconsistent differences between calculated cumulative load and task-based exposure values suggests the presence of random error rather than systematic error which results in insufficiency of the standard correction/approximation factors utilized in task-based assessments to obtain a closer-to-reality exposure calculation.

This finding can be better contextualized in conjunction with empirical studies on muscular activity analyses. For example; Yi et al. [61] measure pulling strengths in a pallet truck handling operation, concluding a role of maximum endurance time (MET) on the development of muscular fatigue in such MMH activities and hence, work/rest allowance in job design. Therefore, a task-based assessment considering the pulling activity as critical in isolation may overlook the cumulative fatigue effects in the absence of MET consideration within the complete work cycle.

**6. Conclusion and future scope**

This paper addresses a crucial research-practice gap by exploring the construct validity of task-based approach to ergonomic assessment given its extensive application in manufacturing industries for leveraging reliability boosting methods of data collection, especially in complex working scenarios such as non-routine work or unfavourable shop-floor conditions. However, theoretical proof to support validity of results for encouraging continued use of this approach is scarce. The paper explains the imperative for reliable and valid ergonomic assessment in the context of non-routine work, the various impediments to the assessment process, corresponding assessment approaches along with the correction factors employed in order to achieve an exposure score true to reality. Hence, based on the utility of task-based assessment method for practical implementation of reliability enhancing data collection techniques such as direct measurement tools, the paper strives to gain insights on the validity of results.

Using the MTMM matrix method, the experimental data suggests a complete lack of convergent validity and a weak discriminant validity which overall indicates inadequate construct validity for the task-based assessment approach. Deeper investigations into the interrelationships between same traits being measured using task-based approach and cumulative load analysis revealed three contributing factors for this inadequacy being overlooked low-salience activities, overlooked unplanned activities additional to the standard process flow, and lack of context and continuity in task execution within task-based assessment approach. Amongst the three observed factors contributing to the lack of construct validity, the overlooking of low-salient activities can be managed through modifications in the sampling strategy enabling a meticulous selection of critical activities for assessment. This makes it a controllable factor to some extent. On the contrary, the remaining two factors are uncontrollable. For instance, overlooking of unplanned activities beyond the predefined process flow, nonetheless come into picture due to the unstructuredness of non-routine work. Similarly, lack of contextual coherence and discontinuity in elemental movements are inevitable when critical activities are to be executed in isolation with the omission of intermediate activities within the process flow.

Findings not only include the lack of convergence in the task-based and cumulative load assessment results, but the implications for practical ergonomic risk estimation explored with paired-t test comparison also suggests a significant difference in the estimated risk and action levels as per RULA criterion which can lead to misinformation regarding ergonomic interventions at workplace and inefficiency in preventing WMSDs. A potential limitation of this study could be the exclusive use of RULA criterion for evaluating risk level discrepancies subsequent to direct measurement data collection. The criterion was selected given its widespread application in risk assessments for most manufacturing tasks. However, it could be worthwhile to explore whether risk and action level discrepancies persist with the utilization of alternative risk estimation criteria such as REBA, NIOSH, EAWS or OWAS.

The interrelationship study also suggested that the deviation observed in the results of task-based approach from actual exposure is random in nature and hence the inefficiency of standard correction factors in the experimental results was justified. In conclusion, these findings suggest a need for cumulative load assessments even while leveraging highly reliable assessment techniques such as direct measurement, to ensure the validity of exposure estimation. Consequently, the need for enhancement of advanced industrial engineering technologies such as the direct measurement techniques ensuring applicability to diverse manufacturing processes such as non-routine, highly dynamic, complex or collaborative work, whilst reducing the prerequisite of work-sampling involved in ergonomic assessments is highlighted. Therefore, future scope indicates exploration of constraints within practical implementations of direct measurement ergonomic assessments in diverse manufacturing processes facilitating cumulative load analysis of ergonomic exposures.

**Disclosure statement**

No potential conflict of interest was reported by the authors.

**References:**

1. Karsh B-T, Moro FBP, Smith MJ. The Efficacy of Workplace Ergonomic Interventions to Control Musculoskeletal Disorders: A Critical Analysis of the Peer-Reviewed Literature. Theor Issues Ergon Sci. 2001;2:23–96. doi: 10.1080/14639220152644533.

2. Genaidy A, Karwowski W, Shoaf C. The Fundamentals of Work System Compatibility Theory: An Integrated Approach to Optimization of Human Performance at Work. Theor Issues Ergon Sci. 2002;3:346–368. doi: 10.1080/14639220210124076.

3. Liu H, Hwang S-L, Liu T-H. Economic assessment of human errors in manufacturing environment. Saf Sci [Internet]. 2009 [cited 2025 May 25];47:170–182. doi: 10.1016/j.ssci.2008.04.006.

4. Chintada A, V U. Improvement of productivity by implementing occupational ergonomics. J Ind Prod Eng. 2021;39:59–72. doi: 10.1080/21681015.2021.1958936.

5. Wodajeneh SN, Azene DK, Berhan E. Impacts of ergonomic risk factors on the well-being and innovation capability of employees in the manufacturing industry. Int J Occup Saf Ergon. 2024;30:412–424. doi: 10.1080/10803548.2024.2313905.

6. Tripathi C, Arora M, Chakrabarti A. Exploring the synergy between ergonomics and productivity in the workplace: an empirical analysis using inertial motion capture. Chakrabarti Suwas Arora M Ed Ind 40 Adv Manuf [Internet]. Bengaluru, India. Singapore: Springer; 2025. Available from: https://doi.org/10.1007/978-981-97-7150-9_24.

7. Tseng M-L, Tran TPT, Ha HM, Bui T-D, Lim MK. Sustainable industrial and operation engineering trends and challenges Toward Industry 4.0: a data driven analysis. J Ind Prod Eng [Internet]. 2021 [cited 2025 May 9];38:581–598. doi: 10.1080/21681015.2021.1950227.

8. Ghobakhloo M, Iranmanesh M, Tseng M-L, Grybauskas A, Stefanini A, Amran A. Behind the definition of Industry 5.0: a systematic review of technologies, principles, components, and values. J Ind Prod Eng [Internet]. 2023 [cited 2025 May 9];40:432–447. doi: 10.1080/21681015.2023.2216701.

9. Ghobakhloo M, Iranmanesh M, Foroughi B, Rejeb A, Nikbin D, Tseng M-L. A practical guide on strategic roadmapping for information and operations technology management: a case study on industry 5.0 transformation. J Ind Prod Eng [Internet]. 2024 [cited 2025 May 9];41:397–421. doi: 10.1080/21681015.2024.2325687.

10. Ghorbani E, Keivanpour S, Sekkay F, Imbeau D. Fuzzy expert system for ergonomic assembly line worker assignment and balancing problem under uncertainty. J Ind Prod Eng [Internet]. 2025 [cited 2025 May 10];42:274–296. doi: 10.1080/21681015.2024.2389963.

11. Yates JW, Karwowski W. Maximum Acceptable Lifting Loads during Seated and Standing Work Positions. Appl Ergon. 1987;18:239–243.

12. Karsh BT. Theories of work-related musculoskeletal disorders: Implications for ergonomic interventions. Theor Issues Ergon Sci. 2006;7:71–88. doi: 10.1080/14639220512331335160.

13. Reiman A. Human Factors and Ergonomics in Manufacturing in the Industry 4.0 Context: A Scoping Review. Technol Soc. 2021;65:101572. doi: 10.1016/j.techsoc.2021.101572.

14. Paquet V, Punnett L, Woskie S. Reliable exposure assessment strategies for physical ergonomics stressors in construction and other non-routinized work. Ergonomics. 2005;Jul;48(9):1200-1219. doi: 10.1080/00140130500197302.

15. Dempsey PG, Mathiassen SE. On the evolution of task-based analysis of manual materials handling, and its applicability in contemporary ergonomics. Appl Ergon. 2006;Jan;37(1):33-43. doi: 10.1016/j.apergo.2004.11.004.

16. Gold JE, Park JS, Punnett L. Work routinization and implications for ergonomic exposure assessment. Ergonomics. 2006;Jan;49(1):12-27. doi: 10.1080/00140130500356643.

17. Buchmeister B, Herzog NV. Advancements in Data Analysis for the Work-Sampling Method. Algorithms. 2024;17:183. doi: 10.3390/a17050183.

18. Menolotto M, Komaris D-S, Tedesco S, O'Flynn B, Walsh M. Motion Capture Technology in Industrial Applications: A Systematic Review. Sensors. 2020;20:5687. doi: 10.3390/s20195687.

19. Stefana E, Marciano F, Rossi D, Cocca P, Tomasoni G. Wearable Devices for Ergonomics: A Systematic Literature Review. 2021. p. 777.

20. Maksimović N, Čabarkapa M, Tanasković M, Randjelović D. Challenging Ergonomics Risks with Smart Wearable Extension Sensors. Electronics. 2022;11:3395. doi: 10.3390/electronics11203395.

21. Dempsey PG. Utilizing criteria for assessing multiple-task manual materials handling jobs. Int J Ind Ergon. 1999;24:405–416.

22. Granata KP, Marras WS, Davis KG. Variation in spinal load and trunk dynamics during repeated lifting exertions. Clin Biomech. 1999;14:367–375.

23. Rajesh R. Manual Material Handling: AClassification Scheme. Procedia Technol. 2016;24:568-575,. doi: 10.1016/j.protcy.2016.05.114.

24. Potvin JR, Ciriello VM, Snook SH, Maynard WS, Brogmus GE. The Liberty Mutual Manual Materials Handling (LM-MMH) Equations. Ergonomics. 2021;64:955–970. doi: 10.1080/00140139.2021.1891297.

25. Waters TR, Putz-Anderson V, Garg A, Fine LJ. Revised NIOSH Equation for the Design and Evaluation of Manual Lifting Tasks. Ergonomics. 1993;36:749–776. doi: 10.1080/00140139308967940.

26. McAtamney L, Corlett EN. RULA: a survey method for the investigation of work-related upper limb disorders. Appl Ergon [Internet]. 1993;24:91-99,. doi: 10.1016/0003-6870(93)90080-S.

27. Buchholz B, Paquet V, Punnett L, Lee D, Moir S. PATH: A work sampling-based approach to ergonomic job analysis for construction and other non-repetitive work. Appl Ergon [Internet]. 1996 [cited 2025 May 7];27:177–187. doi: 10.1016/0003-6870(95)00078-X.

28. Ahmad S, Muzammil M. Revised NIOSH Lifting Equation: A Critical Evaluation. Int J Occup Saf Ergon. 2022;29:358–365. doi: 10.1080/10803548.2022.2049123.

29. Chung AZQ, Shorrock ST. The Research-Practice Relationship in Ergonomics and Human Factors – Surveying and Bridging the Gap. Ergonomics. 2011;54:413–429.

30. Shorrock ST, Williams CA. Human Factors and Ergonomics Methods in Practice: Three Fundamental Constraints. Theor Issues Ergon Sci. 2016;17:468–482. doi: 10.1080/1463922X.2016.1155240.

31. Stanton NA, Young MS. Giving Ergonomics Away? The Application of Ergonomics Methods by Novices. Appl Ergon. 2003;34:479–490.

32. Wilson J. Fundamentals of Ergonomics in Theory and Practice. Appl Ergon. 2000;31:557–567.

33. Stanton NA, Young M. What Price Ergonomics? Nature. 1999;399:197–198.

34. Bolarinwa O. Principles and methods of validity and reliability testing of questionnaires used in social and health science researches. Niger Postgr Med J. 2015;22:195–201. doi: 10.4103/1117-1936.173959.

35. Cronbach LJ, Meehl PE. Construct validity in psychological tests. Psychol Bull. 1955;52:281–302. doi: 10.1037/h0040957.

36. Clark LA, Watson D. Constructing validity: New developments in creating objective measuring instruments. Psychol Assess. 2019;31:1412–1427. doi: 10.1037/pas0000626.

37. Polit DF, Beck CT. The content validity index: Are you sure you know what's being reported? Critique and recommendations. Res Nurs Heal. 2006;29:489–497. doi: 10.1002/nur.20147.

38. Liang Y, Lau PWC, Huang WYJ, Maddison R, Baranowski T. Validity and reliability of questionnaires measuring physical activity self-efficacy, enjoyment, social support among Hong Kong Chinese children. Prev Med Rep. 2014;1:48–52.

39. DePoy E, Gitlin LN. Collecting Data Through Measurement in Experimental-Type Research. In: DePoy E, Gitlin LN, editors. Introd Res [Internet]. 5th ed. Mosby; 2016. p. 227–247. Available from: https://doi.org/10.1016/B978-0-323-26171-5.00017-3.

40. Strauss ME, Smith GT. Construct validity: advances in theory and methodology. Annu Rev Clin Psychol. 2009;5:1–25. doi: 10.1146/annurev.clinpsy.032408.153639.

41. Campbell DT, Fiske DW. Convergent and discriminant validation by themultitrait-multimethod matrix. Psychol Bull. 1959;56:81–105. doi: 10.1037/h0046016.

42. Kenny DA. An empirical application of confirmatory factor analysis to the multitrait-multimethod matrix. J Exp Soc Psychol. 1976;12:247–252. doi: 10.1016/0022-1031(76)90055-x.

43. Kluay-On P, Chaikumarn M. Construct Validity, Internal Consistency and Test-Retest Reliability of Ergonomic Risk Assessment for Musculoskeletal Disorders in Office Workers (ERAMO. Theor Issues Ergon Sci. 2021;23:121–130. doi: 10.1080/1463922X.2021.1922780.

44. Nahm AY, Rao SS, Solis-Galvan LE, Ragu-Nathan TS. The Q-Sort Method: Assessing Reliability And Construct Validity Of Questionnaire Items At A Pre-Testing Stage. J Mod Appl Stat Methods. 2002;1. doi: 10.22237/jmasm/1020255360.

45. Hamdani MR, Valcea S, Buckley MR. The MTMM matrixapproach: Implications for HRM research. Pers Rev. 2016;45:1156–1175. doi: 10.1108/pr-12-2014-0278.

46. Furr R, Bacharach V. Validity: Estimating and evaluating convergent and discriminant validity evidence. Psychom Introd. 2006;191–235.

47. Fiske DW. Construct invalidity comes from method effects.Educational and Psychological Measurement. 1987. p. 285–307.

48. Joshi M, Deshpande V. A systematic review of comparative studies on ergonomic assessment techniques. Int J Ind Ergon. 2019; doi: 10.1016/j.ergon.2019.102865.

49. Comberti L, Demichela M. Customised risk assessment in manufacturing: A step towards the future of occupational safety management. Saf Sci [Internet]. 2022 [cited 2025 May 25];154:105809. doi: 10.1016/j.ssci.2022.105809.

50. Van Dieën JH, Hoozemans MJM, Van Der Beek AJ, Mullender M. Precision of estimates of mean and peak spinal loads in lifting. J Biomech [Internet]. 2002 [cited 2025 May 7];35:979–982. doi: 10.1016/S0021-9290(02)00051-9.

51. Kanis H. Variation in Results of Measurement Repetition of Human Characteristics and Activities. Appl Ergon [Internet]. 1997;28:155–163. doi: 10.1016/S0003-6870(96)00071-3.

52. Kumar S. Cumulative load as a risk factor for back pain. Spine. 1990;15:1311–1316. doi: 10.1097/00007632-199012000-00014.

53. Svendsen SW, Mathiassen SE, Bonde JP. Task based exposure assessment in ergonomic epidemiology: a study of upper arm elevation in the jobs of machinists, car mechanics, and house painters. Occup Environ Med. 2005;62:18–27. doi: 10.1136/oem.2004.015966.

54. Annett J. A Note on the Validity and Reliability of Ergonomics Methods. Theor Issues Ergon Sci. 2002;3:228–232. doi: 10.1080/14639220210124067.

55. Stanton NA. Developing and Validating Theory in Ergonomics Science. Theor Issues Ergon Sci. 2002;3:111–114.

56. Kanis H. Reliability and Validity of Findings in Ergonomics Research. Theor Issues Ergon Sci. 2013;15:1–46. doi: 10.1080/1463922X.2013.802058.

57. Stanton NA. Commentary on the Paper by Heimrich Kanis Entitled 'Reliability and Validity of Findings in Ergonomics Research': Where Is the Methodology in Ergonomics Methods? Theor Issues Ergon Sci. 2013;15:55–61. doi: 10.1080/1463922X.2013.778355.

58. Stanton NA. On the Reliability and Validity of, and Training in, Ergonomics Methods: A Challenge Revisited. Theor Issues Ergon Sci. 2016;17:345–353. doi: 10.1080/1463922X.2015.1117688.

59. Vigoroso L, Caffaro F, Tronci M, Fargnoli M. Ergonomics and design for safety: A scoping review and bibliometric analysis in the industrial engineering literature. Saf Sci [Internet]. 2025 [cited 2025 May 25];185:106799. doi: 10.1016/j.ssci.2025.106799.

60. Wu S-P, Chung H-C. The effect of operating height and holding mode on holding capacity for female subjects. J Ind Prod Eng [Internet]. 2013 [cited 2025 May 7];30:248–255. doi: 10.1080/21681015.2013.818070.

61. Yi C-N, Tang F, Li KW. Muscular strength decrease and maximum endurance time assessment for a simulated truck pulling task. J Ind Prod Eng [Internet]. 2017 [cited 2025 May 9];34:486–493. doi: 10.1080/21681015.2017.1360406.